\documentclass[aps,prb,twocolumn,groupedaddress]{revtex4}
\usepackage[bookmarks]{hyperref}
\usepackage[dvips]{graphicx}
\usepackage{amsmath}
\usepackage{amstext}
\usepackage{graphics}
\usepackage{exscale}
\usepackage{epsfig}
\usepackage{changebar}
\usepackage[T1]{fontenc}
\usepackage{bm}
\usepackage{bbm}
\usepackage{version}%\usepackage{psfrag}

\begin{document}
\title[]{Renormalized perturbation theory and  scaling for an impurity Anderson model} 
\author{K Edwards, A C Hewson and V Pandis} \affiliation{Department of Mathematics, Imperial College, London SW7 2AZ,
 United Kingdom.} 
%% \ead{a.hewson@imperial.ac.uk}
%% \pacs{72.10.F,72.10.A,73.61,11.10.G}
%% 75.20.Hr  kondo
%% 71.10.Ay Fermi liquid
%%  73.63.Kv quantum dots
%%  71.27.+a  heavy fermions

 \begin{abstract}
 We demonstrate the effectiveness of a generalized  renormalized perturbational approach
 to calculate the induced magnetization for the single impurity
 Anderson model with a strong on-site  interaction, using  flow equations for
 renormalized parameters 
 to scale from a weak correlation to a strong correlation regime. We show that, using
 simple approximation schemes in different parameter regimes, remarkably accurate
 results  can be obtained for  all magnetic field values by  comparing the results with
 those from direct numerical renormalization group and Bethe ansatz calculations.

 \end{abstract}

\maketitle

\section{Introduction}
 
The impurity Anderson model in the strong interaction (Kondo) regime
 has served as a classic testing ground for the development of
many-body techniques for application to a wider class of models
for strongly correlated electron systems. This is because exact or very accurate
methods have led to a comprehensive understanding of this model\cite{Hew93b};
notably via the numerical renormalization group (NRG)\cite{Wil75,KWW80a}, Bethe ansatz
method \cite{AFL83,TW83}, and for degenerate models the $1/N$\cite{Kur83} and slave boson methods{\cite{Col84,Rea85}.
Though a direct perturbation expansion in powers of the on-site interaction $U$ works well for low and intermediate 
interaction strengths for the symmetric Anderson
 model\cite{YY75,Yam75a,Yam75b,HSZ87} but it
breaks down in the interesting strong interaction regime where
the model displays all the strong correlation physics of the Kondo model 
(it can be mapped into this model via a Schrieffer-Wolff transformation\cite{SW66}).
The assumption has been, therefore, that the strong correlation low
energy physics can only be obtained using a non-perturbative method.
However,  the low energy fixed point corresponds to a Fermi liquid\cite{Noz74},
which can be  described in terms of   quasiparticles, with the single particle
 excitations in 1-1 correspondence
  of the non-interacting system, together with a limited
number of parameters to describe the inter-quasiparticle interactions. 
As a consequence the perturbation
 theory  can be reorganized  as a renormalized perturbation expansion,
similar to that used in quantum electrodynamics, with parameters
corresponding to renormalized versions of those of the original
 model. The renormalized perturbation theory (RPT) provides asymptotically
exact results for the thermodyamics and dynamics in the low energy
regime \cite{Hew93,Hew06}. 
However, the renormalized parameters
have to be calculated and in the strong correlation regime this would
seem to require a non-perturbative technique such as the NRG\cite{HOM04}.\par

Here we demonstrate that this is not the case. We introduce a generalized
scaling scheme, such that 
we can follow a  set of flow equations for the  renormalized
parameters from a regime of weak to strong electron correlation.
Once these parameters have been determined, there are exact relations
from which  the magnetization, spin and charge susceptibilities
 \cite{Hew93,Hew01,Hew05,HBK06,BH07a} can be calculated.\par
We consider the  Anderson model\cite{And61} in a magnetic field $H$,
\begin{eqnarray}
&&H_{\rm AM}=\sum_{\sigma}\epsilon^{}_{\mathrm{d},\sigma} d^{\dagger}_{\sigma}  
d^{}_{\sigma}+Un_{\mathrm{d},\uparrow}n_{\mathrm{d},\downarrow} \label{ham}\\
&& +\sum_{{ k},\sigma}( V^{}_{ k,\sigma}d^{\dagger}_{\sigma}
c^{}_{{ k},\sigma}+ V_{ k,\sigma}^*c^{\dagger}_{{
k},\sigma}d^{}_{\sigma})+\sum^{}_{{
k},\sigma}\epsilon^{}_{{ k},\sigma}c^{\dagger}_{{ k},\sigma}
c^{}_{{
k},\sigma}, \nonumber
\end{eqnarray}
where $\epsilon_{\mathrm{d},\sigma}=\epsilon_{\rm d}-\sigma g\mu_{\rm B} H/2$,
$\sigma=\pm 1$ ($\uparrow,\downarrow$),
is the energy of the localized  level at an impurity site, $U$ the interaction at this local site, and $V_{k,\sigma}$ the
hybridization matrix element to a band of conduction electrons of spin
$\sigma$ with energy $\epsilon_{k,\sigma}-\sigma g_c\mu_{\rm B} H/2$, where
$g_c$ is the g-factor for the conduction electrons. The resonance width
factor $ \Delta_\sigma(\omega)=\pi\sum_{k}|
V_{k,\sigma}|^2\delta(\omega -\epsilon_{ k,\sigma})$ we take as a constant
$\Delta$, corresponding to a flat wide conduction band, and we use the notation $h=g\mu_{\rm B}H/2$. The model can be
characterized by the set of parameters $(h, \epsilon_d, \Delta_{\uparrow},
\Delta_{\downarrow},U)$,
which we can regard as a defining a point ${\boldsymbol \mu}$  in a configuration space.
\par
The Fourier transform of the retarded single-particle impurity Green's function
has the form,
\begin{equation}
G_{\sigma}(\omega,{\boldsymbol \mu})={1\over
    \omega-\epsilon_{\mathrm{d}\sigma}+i\Delta_\sigma-\Sigma_\sigma(\omega,{\boldsymbol \mu})},
\label{gf}
\end{equation}
 where
$\Sigma_{\sigma}(\omega,{\boldsymbol \mu})$ is the self-energy, which we assume to be non-singular at $\omega=0$ so that
 the low energy fixed point of the system corresponds to a Fermi liquid.
The renormalized parameters\cite{HBK06},   $\tilde\epsilon_{\mathrm{d},\sigma}({\boldsymbol \mu})$ and 
$\tilde\Delta_{\sigma}({\boldsymbol \mu})$, which characterize the quasiparticles of the Fermi liquid are defined by
\begin{equation}
\tilde\epsilon_{\mathrm{d},{\sigma}}({\boldsymbol \mu})=z_{\sigma}(\epsilon_{\mathrm{d},{\sigma}} 
+\Sigma_\sigma(0,{\boldsymbol \mu})),\quad\tilde\Delta_{\sigma}({\boldsymbol \mu}) =z_\sigma\Delta,
\label{ren1}
\end{equation} 
where $z_{\sigma}$ is given by
$z_{\sigma}={1/{(1-\Sigma_{\sigma}'(0,{\boldsymbol \mu}))}}$. 
 It is
useful to introduce a renormalized field $\tilde h({\boldsymbol \mu})$ and average
effective level $\tilde\epsilon_{\mathrm{d}}({\boldsymbol \mu})$ defined by
\begin{equation} \tilde\epsilon_{\mathrm{d}}({\boldsymbol \mu})=
  {1\over 2}\sum_\sigma\tilde\epsilon_{\mathrm{d},{\sigma}}({\boldsymbol \mu}),\quad
\tilde h({\boldsymbol \mu})=-{1\over
  2}\sum_\sigma \sigma\tilde\epsilon_{\mathrm{d},{\sigma}}({\boldsymbol \mu}).
\end{equation}
 We can then define
a quasiparticle Green's function via $G_{\sigma}(\omega,{\boldsymbol \mu})=z_\sigma\tilde
G_{\sigma}(\omega,{\boldsymbol \mu})$
such that 
\begin{equation}
\tilde G_{\sigma}(\omega,{\boldsymbol \mu})={1\over
    \omega-\tilde\epsilon_d({\boldsymbol \mu})+\sigma \tilde h({\boldsymbol
    \mu})+i\tilde\Delta_\sigma({\boldsymbol\mu})-\tilde\Sigma_\sigma(\omega,{\boldsymbol \mu})},
\label{qpgf}
\end{equation}
where $\tilde\Sigma_\sigma(\omega,{\boldsymbol \mu})$ is the renormalized self-energy.
The effective interaction between the quasiparticles  $\tilde U({\boldsymbol\mu})$ can be  defined in terms
of the local total 4-vertex
$\Gamma^{(4)}_{\uparrow\downarrow}(\omega_1,\omega_2,\omega_3,\omega_4:{\boldsymbol \mu})$ at zero frequency,
  \begin{equation} 
\tilde U({\boldsymbol
  \mu})=z_{\uparrow}z_{\downarrow}\Gamma^{(4)}_{\uparrow\downarrow}(0,0,0,0:{\boldsymbol
  \mu}).\label{ren2}\end{equation}
The  Lagrangian corresponding to the bare Anderson  ${\cal L}_{\rm
AM}(\boldsymbol\mu)$ can be rewritten in  the form, 
\begin{equation}
{\cal L}_{\rm AM}({\boldsymbol\mu})={ \cal L}_{\rm AM}(\tilde{\boldsymbol\mu})+ {\cal L}_{\rm
  ct}({\boldsymbol\lambda}),\label{rlag}
\end{equation}
where $\boldsymbol\lambda$ are the coefficients in the remaining  or
counterterm part of the
Lagrangian. They can be determined from the renormalization conditions, 
\begin{equation}\tilde\Sigma(0,\boldsymbol\mu)=0,\quad \tilde\Sigma'(0,\boldsymbol\mu)=0,\quad
  \tilde\Gamma_{\uparrow,\downarrow}(0,0,0,0:\boldsymbol\mu)=\tilde U(\boldsymbol\mu). \label{rcond}\end{equation}
These ensure that there is no overcounting of renormalization effects
which have already been included by using the renormalized parameters
instead of the bare ones.\par
Exact results  can be derived for a number of physical quantities in the Fermi
 liquid regime in terms of the renormalized
 parameters\cite{Hew93}. Calculations can be carried out for the spectral functions and behavior
beyond the Fermi liquid regime using renormalized perturbation theory (RPT)
in terms of the renormalized parameters as no approximation is made in setting
up the expansion \cite{Hew01,Hew06}.\par

The renormalized parameters can be regarded as defining an alternative
specification of the model, $\tilde{\boldsymbol \mu}=(\tilde h({\boldsymbol\mu}),\tilde
\epsilon_d({\boldsymbol\mu}),\tilde \Delta_\uparrow({\boldsymbol\mu}),\tilde \Delta_\downarrow({\boldsymbol\mu}),\tilde U({\boldsymbol\mu}))$. 
Here we introduce the idea of deriving  a scaling relation linking the renormalized
parameters for a model specified by the parameter set ${\boldsymbol \mu}_1$ to one
specified by the parameter set ${\boldsymbol\mu}_2$, $\tilde{\boldsymbol \mu}_1({\boldsymbol \mu}_1)
\to \tilde{\boldsymbol \mu}_2({\boldsymbol\mu}_2)$.
We consider a renormalized perturbation expansion for
$G_{\sigma}(\omega,{\boldsymbol \mu}_2)$, in which  we use the non-interacting
quasiparticle propagator for the system with ${\boldsymbol \mu}={\boldsymbol
  \mu}_1$ instead of that for the system with ${\boldsymbol \mu}={\boldsymbol
  \mu}_2$. Expressing this Green's function 
 in the form,
\begin{equation}
G_{\sigma}(\omega,{\boldsymbol \mu}_2)={z_\sigma({\boldsymbol\mu}_1)\over
    (\tilde G_\sigma^{(0)}(\omega,{\boldsymbol
    \mu}_1))^{-1}-\tilde\Sigma_\sigma(\omega,{\boldsymbol
    \mu}_1,{\boldsymbol \mu}_2-{\boldsymbol \mu}_1)},
\label{nqpgf}
\end{equation}
effectively defines the self-energy $\tilde\Sigma_\sigma(\omega,{\boldsymbol
    \mu}_1,{\boldsymbol \mu}_2-{\boldsymbol \mu}_1)$. The corresponding
Larangian takes the form,
\begin{equation}
{\cal L}_{\rm AM}(\boldsymbol\mu_2)={ \cal L}_{\rm AM}(\tilde{\boldsymbol\mu}_1)+ {\cal L}_{\rm ct}({\boldsymbol\lambda}_1)+{\cal L}_{\rm
  ex}({\boldsymbol\mu}_2-{\boldsymbol\mu}_1)
,\label{rlag2}
\end{equation}
where  there is an additional term ${\cal L}_{\rm
  ex}({\boldsymbol\mu}_2-{\boldsymbol\mu}_1)$ due to the difference between
  the models with different parameter sets. The counterterms in  ${\cal L}_{\rm ct}({\boldsymbol\lambda}_1)$
  are required to satisfy the conditions (\ref{rcond}) for ${\boldsymbol\mu}=
{\boldsymbol\mu}_1$.
\par

By equating the inverse of $G_{\sigma}(\omega,{\boldsymbol \mu}_2)$ from
    Eqn. (\ref{nqpgf}) with that derived from  Eqn. (\ref{qpgf})
and its derivative at $\omega =0$, we can relate the renormalized
parameters at ${\boldsymbol \mu}_1$ to those for ${\boldsymbol \mu}_2$.
For ${\boldsymbol \mu}_1={\boldsymbol \mu}$, ${\boldsymbol \mu}_2={\boldsymbol \mu}+\delta{\boldsymbol \mu}$, these equations,
take the form,
\begin{equation} \tilde\Delta_\sigma({\boldsymbol \mu}+\delta{\boldsymbol \mu})=\bar z_{\sigma}(
{\boldsymbol \mu},\delta{\boldsymbol \mu}) \tilde\Delta_\sigma({\boldsymbol \mu}), \label{scale_del}
\end{equation}
where $\bar z_{\sigma}(
{\boldsymbol \mu},\delta{\boldsymbol \mu})=
(1-\tilde\Sigma'_\sigma(0,{\boldsymbol \mu},\delta{\boldsymbol \mu}))^{-1}$,
 and
\begin{equation} \tilde\epsilon_d({\boldsymbol \mu}+\delta{\boldsymbol \mu})
-\sigma\tilde h({\boldsymbol \mu}+\delta{\boldsymbol \mu})
=\bar z_{\sigma}( \tilde\epsilon_d({\boldsymbol \mu})
-\sigma\tilde h({\boldsymbol \mu})+\tilde\Sigma_\sigma(0,{\boldsymbol
  \mu},\delta{\boldsymbol \mu})).
\label{scale}
\end{equation}
The self-energy $\tilde\Sigma_\sigma(\omega,{\boldsymbol \mu},\delta{\boldsymbol \mu})$
and its derivative are zero  at $\omega=0$  for $\delta{\boldsymbol \mu}=0$ , so we take them to lowest  order in
 $\delta{\boldsymbol \mu}$, which in general will be first order in  $\delta{\boldsymbol \mu}$. We write them as
 $\tilde\Sigma_\sigma(0,{\boldsymbol \mu},\delta{\boldsymbol
   \mu})=\alpha_{1,\sigma}({\boldsymbol \mu},\delta{\boldsymbol \mu})$ and
$\tilde\Sigma'_\sigma(0,{\boldsymbol \mu},\delta{\boldsymbol
   \mu})=\alpha_{2,\sigma}({\boldsymbol \mu},\delta{\boldsymbol \mu})$.
Working to lowest order in  $\delta{\boldsymbol \mu}$ we can separate (\ref{scale})
into two equations,
\begin{eqnarray}\tilde\epsilon_d({\boldsymbol\mu}+\delta
  {\boldsymbol\mu})-\tilde\epsilon_d({\boldsymbol\mu})&=&\\
\tilde\epsilon_d({\boldsymbol\mu})\alpha_{2,+}({\boldsymbol \mu},\delta{\boldsymbol \mu})&-&\tilde h({\boldsymbol\mu})\alpha_{2,-}({\boldsymbol \mu},\delta{\boldsymbol \mu})+
\alpha_{1,+}({\boldsymbol \mu},\delta{\boldsymbol \mu}).\nonumber
\label{scale_ed}\end{eqnarray}
\begin{eqnarray}\tilde h({\boldsymbol\mu}+\delta
  {\boldsymbol\mu})-\tilde h({\boldsymbol\mu})&=&\\
\tilde h({\boldsymbol\mu})\alpha_{2,+}({\boldsymbol \mu},\delta{\boldsymbol \mu})&-&\tilde \epsilon_d({\boldsymbol\mu})\alpha_{2,-}({\boldsymbol \mu},\delta{\boldsymbol \mu})-
\alpha_{1,-}({\boldsymbol \mu},\delta{\boldsymbol \mu}).\nonumber
\label{scale_h}\end{eqnarray}
where 
\begin{equation}\alpha_{i,\pm}({\boldsymbol \mu},\delta{\boldsymbol \mu})=0.5(
\alpha_{i,\uparrow}({\boldsymbol \mu},\delta{\boldsymbol \mu})\pm \alpha_{i,\downarrow}({\boldsymbol \mu},\delta{\boldsymbol \mu})).\end{equation}

To apply these scaling equations to calculate the renormalized parameters
in a strong correlation regime two conditions need to be satisfied: (i) 
we need a known set of renormalized parameters corresponding to ${\boldsymbol \mu}_1$,
${\boldsymbol \mu}_1\to \tilde{\boldsymbol \mu}_1$, and (ii) an accurate method of
calculating the renormalized self-energy
$\tilde\Sigma_\sigma(\omega,{\boldsymbol\mu},\delta{\boldsymbol\mu})$ in the low frequency range,
ie. its value and derivative as $\omega \to 0$. The results should be
independent of the particular trajectory chosen for generating ${\boldsymbol \mu
}_2$  from ${\boldsymbol \mu
}_1$. Note that in contrast to the Wilson renormalization approach,
where states are integrated out,  one can define an inverse transformation
 $\tilde{\boldsymbol \mu}_2({\boldsymbol \mu}_2)
\to \tilde{\boldsymbol \mu}_1({\boldsymbol \mu}_1)$.
Condition (i) can be satisfied  by choosing ${\boldsymbol \mu}_1$ in a weakly
correlated regime where perturbation theory can be applied. The more difficult
step is to find an approximation to satisfy condition (ii). \par
\par
In particular cases one can use additional information derived from relevant Ward
identities. We consider the case where we change only the parameter
$\mu_1=h$. Using the Friedel sum rule\cite{Lan66} and a Ward identity\cite{Yam75a,Yam75b} we derive the
result,
\begin{equation}
\alpha_{1,\sigma}(h,\delta
  h)=-\sigma(1+\tilde U(h)\tilde\rho_{-\sigma}(0,h))\delta h+{\rm O}((\delta
h)^2),\label{w1}\end{equation}
where
\begin{equation}
\tilde\rho(\omega,{\boldsymbol\mu})
 ={1\over \pi}{\tilde\Delta({\boldsymbol\mu})\over (\omega-\epsilon_d({\boldsymbol\mu})-\tilde h({\boldsymbol\mu})^2
+\tilde\Delta^2({\boldsymbol\mu})}
,\label{tilrho}\end{equation}
 The details of the derivation are given in the Appendix A.
We can  use this result in the flow equations. For $h=0$, $\tilde\epsilon_d(\delta h)$
and $\alpha_{2,\sigma}(0,0,\delta h)$ are of order $(\delta h)^2$, and from the
  scaling equations we find $\tilde
  h(h)/h\to R$ where $R=1+\tilde U(0)\tilde\rho(0,0)$, is the Wilson ratio,
which is an exact result.\par 
We can derive similar relations in the case where we change
  ${\mu_2}=\epsilon_d$,
\begin{equation}
\alpha_{1,\sigma}(\epsilon_d,\delta
  \epsilon_d)=(1-\tilde U(\epsilon_d)\tilde\rho_{-\sigma}(0,\epsilon_d))\delta \epsilon_d+{\rm O}((\delta
\epsilon_d)^2),\label{w2}\end{equation}
where again details of the derivation are given in the Appendix A. 
 If we change both
  $\epsilon_d$ and $h$, then these two results can be combined in the scaling equation 
to first order in both $\delta\epsilon_d$ and
  $\delta h$.
We use  these results in the scaling equations to show that suitable
  approximations can be derived 
to satisfy both conditions (i) and (ii) in particular parameter regimes.\par
\section{Scaling with Magnetic Field}

We show results first of all for the particle-hole symmetric model. 
In this case  we can satisfy condition (i) by considering the system
in a very large magnetic field so as to suppress the spin fluctuations
that lead to the strong correlation effects in this regime. 
 We then test the accuracy of the approximation used in step (ii) by calculating the induced $m(h)$ magnetization at
$T=0$, which is given in terms of the  renormalized parameters
$\tilde\epsilon_{d,\sigma}$ and $\tilde\Delta_\sigma$  by 
\begin{equation}
m(h)=-{1\over 2\pi}\,\sum_\sigma \sigma\,{\rm tan}^{-1}
\left({\tilde\epsilon_{d,\sigma}\over\tilde\Delta_\sigma}\right),
\label{magqp}
\end{equation}
 and  compare the results with the corresponding  NRG and Bethe ansatz results. For particle-hole symmetry
the formula for the magnetization given in Eqn. (\ref{magqp}) simplifies as $\tilde\epsilon_{d,\sigma}=-\sigma\tilde h$ and $\tilde\Delta_\sigma$
is independent of $\sigma$.
 \par

 \begin{figure}[!htbp]
    \begin{center}
      \includegraphics[width=0.2\textwidth]{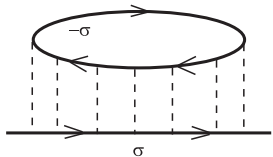}
      \caption{(Color online) A diagram for the (renormalized) self-energy
 involving a repeated scattering in the particle-hole channel, corresponding
 to  scattering with a spin flip. In the initial calculation  in very large magnetic fields, the dashed line represents   the 'bare'   interaction $U$ and  the full lines are the propagators
 calculated in mean field theory in an applied magnetic field $h$. In the renormalized perturbation theory
 the interaction parameter is $\tilde U^t_{ph}$ and the propagators are for
 the quasiparticles in an effective field $\tilde h$ and include the mean
 field insertions. 
 }
      \label{self_ph}
    \end{center}
  \end{figure}
\begin{figure}[!htbp]
    \begin{center}
      \includegraphics[width=0.5\textwidth]{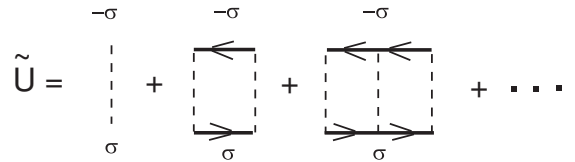}
      \caption{(Color online)  The diagrams, at zero external frequency $\omega=0$,  used to calculate  the effective interaction 
 vertex $\tilde U$ in the particle-hole symmetric regime in an applied
 magnetic field. In the inital calculation the propagators correspond to
mean field theory and the interaction parameter (dashed line) to $U$.
 In the RPT calculations, the  propagators correspond to free quasiparticles
 and the interaction parameter (dashed line) is $\tilde U^t_{ph}$.
 }
      \label{ph_vertex}
    \end{center}
  \end{figure}

Accurate initial renormalized parameters in the extreme large field regime
$h>U$  can be calculated simply  using the original Anderson mean field
theory \cite{And61}. These  can then be improved upon using the set of RPA spin
flip scattering terms to calculate the self-energy 
and the corresponding value of $\tilde U$.
This is illustrated in  Fig. \ref{self_ph} and Fig. \ref{ph_vertex}
where the propagators correspond to those for the ``bare'' electrons with
mean field insertions and the interaction parameter represented by a dashed
line is the bare interaction  $U$.\par

To extend the results to lower magnetic field values we use the scaling
approach
described in the previous section and reduce the applied field to $h-\delta h$
but use the renormalized parameter at the field value $h$. We then need
a suitable  approximation to calculate the renormalized self-energy 
$\tilde\Sigma_\sigma(\omega,h,\delta h)$.
 We have shown that accurate results can be obtained
for the renormalized parameters by using the same set of diagrams
for the self-energy and the vertex $\tilde U$ as in  Fig. \ref{self_ph} and
Fig. \ref{ph_vertex}, where now the propagators correspond to the free
quasiparticles \cite{EH11}. The counterterms have to be taken into account and
as a consequence the interaction parameter is modified to  $\tilde U
-\lambda_3$, which we denote by  $\tilde U^t_{ph}$, due to the counterterm
$\lambda_3$. Given the starting values
of $\tilde h(h)$ and $\tilde\Delta(h)$ (for particle-hole
  symmetry $\tilde\epsilon_d=0$) the corresponding value of  $\tilde U^t_{ph}(h)$ can be deduced
  from 
\begin{equation}
{m(h)\over h}={1\over \pi h}{\rm tan}^{-1}
\left({\tilde h(h)
\over\tilde\Delta(h)}\right)={\tilde\Pi_{ph}(0,h)\over 1-\tilde U^t_{ph}\tilde\Pi_{ph}(0,h)},
\label{uph}
\end{equation}
where $\tilde\Pi_{ph}(\omega,h)$ is the transverse dynamic spin susceptibility
due to the non-interacting quasiparticles\cite{Hew06}. The full expression for
$\tilde\Pi_{ph}(\omega,h)$ is given in the Appendix B, and its value at
$\omega=0$ is given by
\begin{equation}
\tilde\Pi_{ph}(0,h)={1\over \pi \tilde h}{\rm tan}^{-1}
\left({\tilde h(h)
\over\tilde\Delta(h)}\right).
\label{piph}
\end{equation}
 Eqns. (\ref{uph}) and (\ref{piph}) imply that $\tilde h(h)$ takes a
 generalized mean field form  $\tilde h(h)=h+\tilde U^t_{ph}(h)m(h)$.\par

The new parameters $\tilde h$ and $\tilde\Delta$ are then calculated
at the reduced field $h-\delta h$ from the scaling equations (\ref{scale_del}) and
(\ref{scale_h}) using Eqn. (\ref{w1}) with $\tilde U(h)$ given by 
\begin{equation}
\tilde U(h)={\tilde U^t_{ph}\over 1-\tilde U^t_{ph}\tilde\Pi_{ph}(0,h)},
\label{tilu}
\end{equation}
corresponding to the diagrams in Fig. \ref{ph_vertex}.
Once  $\tilde h$ and $\tilde\Delta$ at $h-\delta h$ have been calculated the corresponding
values of $U^t_{ph}$ and  $\tilde U$
can be deduced from Eqns. (\ref{uph}) and  (\ref{tilu}) and the
process repeated to reduce the magnetic field in a sequence of steps to zero.\par

 Results for the induced
magnetization  for the case, $U/\pi\Delta=3$, $\pi\Delta=0.1$ and a conduction
band halfwidth $D=1$, are shown  in Figs. \ref{mag1} and  \ref{mag2},
together with those from a direct
NRG calculation. The calculated  Wilson
ratio, $R=1+\tilde U(0)/\pi\tilde\Delta(0)=1.99$, showing that to an excellent
approximation $\tilde U(0)=\pi\tilde\Delta(0)$ so that there is only a single
energy scale corresponding to the Kondo temperature $T_{\rm K}$.   The Kondo temperature $T_{\rm K}$ is defined in terms of  the zero field 
spin susceptibility  $\chi_s$ at $T=0$ via $\chi_s= (g\mu_{\rm
  B})^2S(S+1)/3T_{\rm K}$ with $S=1/2$. The value of $T_{\rm K}$ is 0.00800 which is very close
to the Bethe ansatz result for these parameters  0.00805\cite{HZ85}. The results in Fig.
\ref{mag1} are plotted as a function of ${\rm ln}(h/T_{\rm K})$ over the whole
range of magnetic field values to the saturated value $m(h)\to 1/2$. It can be
seen that there is excellent agreement between the results deduced from the 
RPT scaling equations and the ones determined from a direct NRG
calculation\cite{HBK06}. In Fig. \ref{mag2} we plot and compare the NRG results
as a function of  $h/T_{\rm K}$ up to a value $h=4T_{\rm K}$, and also 
 with Bethe ansatz results \cite{AFL83,TW83} for the s-d model over this range.  It can be seen
again that on this scale there is still excellent agreement between the
results.  
 For larger values of the magnetic field  $h>8T_{\rm K}$ for $U/\pi\Delta=3$
the approach to saturation is different for the s-d and Anderson models because
in this range the impurity
charge excitations play a role for the Anderson model  which are not present in the s-d model\cite{HBK06}.
\par 
In earlier results\cite{EH11}  using this approach we have given more extensive
comparisons with Bethe ansatz and NRG results 
including calculations of the self-energy and
dynamical spin susceptibilities.
We note a scaling approach using a magnetic field using  the functional
renormalization group method (fRG) has recently been developed by Streib,
Isidori and Kopietz \cite{SIKpre12} for the strong correlation regime
for  the particle-hole symmetric model.\par

\begin{figure}[!htbp]
   \begin{center}
     \includegraphics[width=0.33\textwidth]{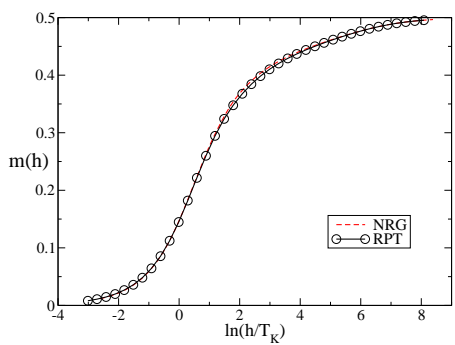}
     \caption{(Color online) The  magnetization $m(h)$  as a function of a function of ${\rm ln}(h/T_{\rm K})$
       calculated from the RG flow in the RPT (dashed line, circles) for the symmetric model with
       $U/\pi\Delta=3$, $\pi\Delta=0,1$ and $T_{\rm K}=0.008$ compared with the corresponding  results of
a direct NRG calculation (dashed line).  }
     \label{mag1}
   \end{center}
 \end{figure}
\bigskip
\begin{figure}[!htbp]
   \begin{center}
     \includegraphics[width=0.33\textwidth]{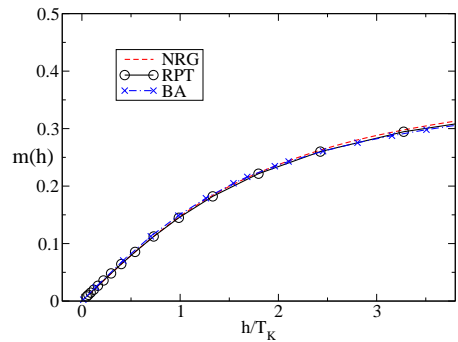}
     \caption{(Color online) The magnetization $m(h)$ as a function of
       $h/T_{\rm K}$ calculated from the RG flow in the
       RPT  (full line, circles) for
       the same parameter set as in Fig. \ref{mag1} compared with the corresponding  results of
a direct NRG calculation (dashed line)
and Bethe ansatz results \cite{AFL83,TW83}(dot-dashed, crosses)  
for the s-d model.}
     \label{mag2}
   \end{center}
 \end{figure}
\section{Scaling relative to  the chemical potential}
Well away from particle-hole symmetry with either a low or
high density of particles or holes, we have another regime  where the correlation effects
are weak. Here we can also apply mean field theory to calculate the 
initial values for the renormalized parameters, and then extend  the calculation of the
self-energy using standard prturbation theory in $U$ to take into account the diagrams
with repeated particle-particle scattering (see Fig. \ref{self_pp}). We
consider this situation initally without an applied magnetic field and introduce a magnetic field later.
 This regime was first
studied by  Schrieffer and Mattis \cite{SM65}. They calculated  an  effective
interaction $U_{\rm eff}$ from an expansion in powers of $U$ corresponding to the diagrams shown in Fig. \ref{pp_vertex},
\begin{equation}
U_{\rm eff}={U\over 1+{U\over \pi \bar\epsilon_d}{\rm tan}^{-1}\left({
      \bar\epsilon_d\over \Delta}\right)},
\label{MS}
\end{equation}
where $\bar\epsilon_d$ is the renormalized level calculated in mean field
theory. 
 The value of  $U_{\rm eff}$ can be identified as $\tilde U$ and is in good agreement with the values deduced from the NRG in
the regime well away from particle-hole symmetry.  To scale from this regime  to a more strongly correlated one
we decrease the value of $|\epsilon_d|$, which is equivalent to changing
$\epsilon_d$  relative to the chemical potential.
We use, as an approximation for the renormalized self-energy in
the scaling equations for part (ii) of the calculation, the same set of diagrams with repeated particle-particle
scattering in the RPT but using quasiparticle propagators  with a renormalized interaction in this channel $\tilde
U_{pp}$. The initial value of $\tilde U-\lambda_3=\tilde U_{pp}$ is calculated
from  
\begin{eqnarray}
 {1- n_{d,\uparrow}-n_{d,\downarrow}
\over 2E_d}={1\over \pi E_d}\,{\rm
    tan}^{-1}\left({\tilde\epsilon_{d}\over \tilde\Delta}\right)\nonumber\\
={{\rm tan}^{-1}\left({
      \tilde\epsilon_d\over \tilde\Delta}\right)\over \pi\tilde\epsilon_d
+\tilde U_{pp}\,{\rm tan}^{-1}\left({
      \tilde\epsilon_d\over \tilde\Delta}\right)},
\label{U_pp}
\end{eqnarray}
 and the initial value of $\tilde U$ from
(\ref{MS}). We have introduced the notation $E_d=\epsilon_d+U/2$, so that
$E_d=0$ corresponds to the model with particle-hole symmetry.
\par
Once the values of $\tilde\epsilon_d$ and $\tilde \Delta$ have been calculated
from the scaling equations at $\epsilon_d-\delta\epsilon_d$,  new  values
of $\tilde U_{pp}$ can be calculated from Eqn. (\ref{U_pp}) at
$\epsilon_d-\delta\epsilon_d$ and the new values of $\tilde U$ deduced from
\begin{equation}
\tilde U={\tilde U_{pp}\over 1+{\tilde U_{pp}\over \pi\tilde\epsilon_d}\,{\rm tan}^{-1}\left({
      \tilde\epsilon_d\over \tilde\Delta}\right)},
 \label{MS2}
\end{equation} 
corresponding to the diagrams in Fig. \ref{pp_vertex}. Note in this case, in scaling with $\epsilon_d$ in the flow equations, we are changing
the model itself rather than an external field.
\par
To test the range of validity of this approximation we calculate the
renormalized parameters from the flow equations and compare them the
corresponding values deduced from an  NRG analysis of the low energy fixed
point for the case $U/\pi\Delta=3$, $\pi\Delta=0.1$ over a range of values
of $\epsilon_d$.   The approximation based on this set of diagrams works
well in the range $|E_d|>U/2$.  The results are shown in
Fig. \ref{params}. Corresponding results for the static spin susceptibility
due to the impurity, which can be evaluated from the renormalized parameters
using the exact result,
\begin{equation}
\chi_s={(g\mu_{\rm B})^2\over 2}\tilde\rho(0,\epsilon_d)(1+\tilde U(\epsilon_d)
\tilde\rho(0,\epsilon_d)),\end{equation}
are shown in Fig. \ref{chi_s}. The values of $\chi_s$ are shown on a logarithmic
scale and are in good agreement with the NRG results.
\par

To calculate the induced magnetization $m(h)$ in a magnetic field in this regime
the formulae become more complicated. The equation for  $\tilde U_{pp}$
generalizes to
\begin{equation}
{ n_{d,\uparrow}+n_{d,\downarrow}
-1\over 2E_d}={\tilde\Pi_{pp}(0)\over 1-\tilde U_{pp}\tilde\Pi_{pp}(0)}
\end{equation}
and for $\tilde U$,
\begin{equation}
\tilde U={\tilde U_{pp}\over 1-\tilde U_{pp}\tilde\Pi_{pp}(0)},
 \label{MS3}
\end{equation}
where
\begin{eqnarray}
\tilde\Pi_{pp}(0)
&=&-{1\over \pi}{2\tilde\epsilon_d\over
  4\tilde\epsilon_d^2+(\tilde\Delta_\uparrow -\tilde\Delta_\downarrow)^2}
\sum_\sigma{\rm tan}^{-1}\left({\tilde\epsilon_{d,\sigma}\over
    \tilde\Delta_\sigma}\right)\nonumber\\ 
&+& {1\over 2\pi}{(\tilde\Delta_\uparrow-\tilde\Delta_\downarrow)
\over
  4\tilde\epsilon_d^2+(\tilde\Delta_\uparrow -\tilde\Delta_\downarrow)^2}\,
{\rm ln}\left[{\tilde\epsilon_{d,\uparrow}^2+\tilde\Delta_\uparrow^2
\over \tilde\epsilon_{d,\downarrow}^2+\tilde\Delta_\downarrow^2}\right]
\end{eqnarray}
\begin{figure}[!htbp]
    \begin{center}
      \includegraphics[width=0.2\textwidth]{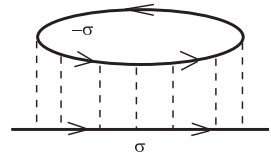}
      \caption{(Color online) A diagram for the (renormalized) self-energy
 involving a repeated scattering in the particle-particle channel. In the
 initial calculation in the low  density regime of either particles or holes the dashed line representing  an interaction
 vertex is the 'bare'   interaction $U$ and  the full lines are the propagators
 calculated in mean field theory. In the renormalized perturbation theory
 the interaction parameter is $\tilde U_{pp}$ and the propagators are for
 the quasiparticles in an effective field $\tilde\epsilon_d$ and includes the mean
 field insertions.
 }
      \label{self_pp}
    \end{center}
  \end{figure}

\begin{figure}[!htbp]
    \begin{center}
      \includegraphics[width=0.5\textwidth]{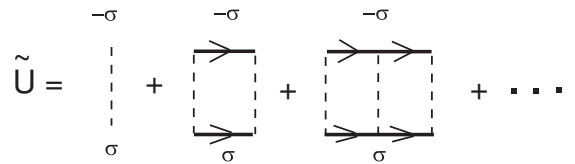}
      \caption{(Color online)  The diagrams, at zero external frequency $\omega=0$,  used to calculate  the effective interaction 
 vertex $\tilde U$ in the  regime  $|E_d|>U/2$.
 In the initial calculation the  propagators correspond to mean field theory
and the interaction parameter (dashed line) is $\tilde U$.
In the RPT calculation the propagators correspond to  free quasiparticles, include the mean
 field insertions, and 
 the interaction parameter (dashed line) is $\tilde U_{pp}$.
}
      \label{pp_vertex}
    \end{center}
  \end{figure}
 
\begin{figure}
   \begin{center}
    \includegraphics[width=0.41\textwidth]{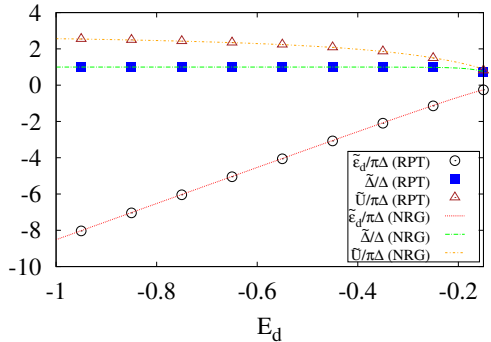}
     \caption{(Color online) The renormalized parameters $\tilde\epsilon_d/\pi\Delta$, $\tilde\Delta/\Delta$ and $\tilde U/\pi\Delta$ as a function of
       $E_d$ calculated from the RG flow in the
       RPT  for the
        model with $U/\pi\Delta=3$, $\pi\Delta=0.1$,   compared with the corresponding  results of
an NRG calculation.
}
     \label{params}
   \end{center}
 \end{figure}

\begin{figure}
   \begin{center}
    \includegraphics[width=0.4\textwidth]{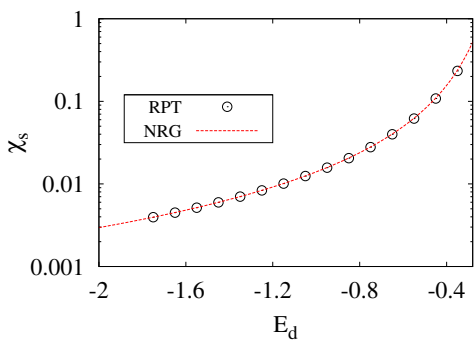}
     \caption{(Color online) The static susceptibility (on a log scale) in the absence of a magnetic field  as a function of
   $E_d$    calculated from the RG flow in the
       RPT  for the
        model with $U/\pi\Delta=3$, $\pi\Delta=0.1$  compared with the corresponding  results of
a direct NRG calculation.}
     \label{chi_s}
   \end{center}
 \end{figure}

The flow equations are solved as a function of $\epsilon_d$ for a given
magnetic field value $h$, and the renormalized parameters used to deduce
the induced magnetization using Eqn. (\ref{magqp}).
 The  results  for $m(h)$ as a 
function of ${\rm ln}(h/T_{\rm K})$ for the case
$U/\pi\Delta=3$, $\pi\Delta=0.1$ and $E_d=-0.15$ are shown in Fig \ref{mag3}. This is a
significantly
correlated regime with $\tilde U/U=0.29$ and $\tilde U\tilde\rho(0)=0.54$.
the value of $T_{\rm K}=0.052$ and is defined in terms of the $T=0$
susceptibility
as earlier. 
 A comparison is made with the corresponding
results from a direct NRG calculation. The two sets of results can be seen to
be in excellent agreement over the full range of magnetic field values.
In Fig. \ref{mag4} we compare the two sets of results for smaller values of
the magnetic field, over a more physically accessible range, as a function of
$h/T_{\rm k}$.\par
As already noted, the approximation used for the calculation of the renormalized
parameters, based on the particle-particle scattering diagrams, breaks down
at the point $|E_d|\sim U/2$.  The difficulty in
extending the calculation beyond this point to the particle-hole symmetric
point  $|E_d|=0$ is that, in the strong correlation regime $U/\pi\Delta\gg 1$, the  effective 
interaction in the particle-particle channel $\tilde U_{pp}$
becomes very large so as to suppress the charge
fluctuations.  
In the strong correlation regime near half-filling $\tilde U\tilde\rho(0)\to
1$ so that the impurity charge susceptibility, which is proportional to $1-\tilde U\tilde\rho(0)$, is effectively zero.
  It can be seen from the
scaling Eqn. (\ref{scale_ed}) and Eqn. (\ref{w2}) that when  $\tilde U\tilde\rho(0)=1$  the first order
term in $\delta\epsilon_d$ is zero. Near particle-hole symmetry for strong
correlation, the main effect is simply a change in the Kondo temperature
$T_{\rm K}$, as   $\tilde
U\tilde\rho(0)=1$ is maintained in this regime,  and $T_{\rm K}(E_d)-T_{\rm K}(0)\propto E_d^2$.\par
An alternative strategy to calculate the renormalized parameters in the strong
coupling regime close to particle-hole symmetry would be to scale from
large to small magnetic fields, based on the particle-hole scattering diagrams
 in Fig. \ref{self_ph}
and  Fig. \ref{ph_vertex}. Away from particle-hole
symmetry with moderate to large magnetic fields there is a difference between  $\tilde\Delta_\uparrow(h)$ and
$\tilde\Delta_\downarrow(h)$ which should go to zero as the magnetic field is
reduced, However, in the numerical calculations a small difference persists as $h\to 0$,
so an improved or alternative approximation is required to obtain completely
satisfactory results in this parameter regime. This is currently being investigated.  

\par\begin{figure}[!htbp]
   \begin{center}
    \includegraphics[width=0.4\textwidth]{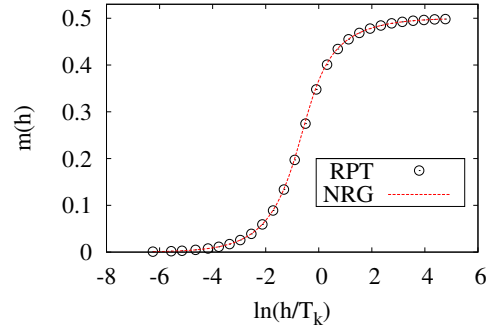}
     \caption{(Color online) The magnetization $m(h)$ as a function of
       ${\rm ln}(h/T_{\rm K})$ calculated from the RG flow in the
       RPT (full line, circles) for the
        model with $U/\pi\Delta=3$, $\pi\Delta=0,1$, $E_d=-0.15$ and $T_{\rm K}=0.052$,  compared with the corresponding  results of
a direct NRG calculation (dashed line)}
     \label{mag3}
   \end{center}
 \end{figure}

\begin{figure}[!htbp]
   \begin{center}
     \includegraphics[width=0.36\textwidth]{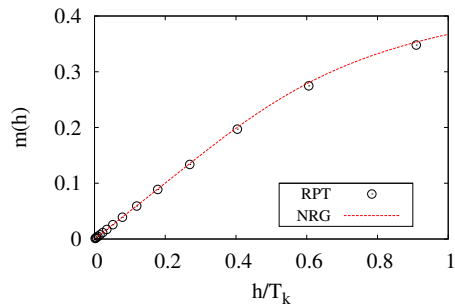}
     \caption{(Color online) The  magnetization $m(h)$ as a function of
       $h/T_{\rm K}$ calculated from the RG flow in the
       RPT (full line, circles) for the
       same parameter set as in Fig. \ref{mag3} compared with the
       corresponding results of
a direct NRG calculation (dashed line)}
     \label{mag4}
   \end{center}
 \end{figure}

\section{Conclusions}

We have shown how, based on a relatively simple set of diagrams
within a renormalized perturbation theory (RPT), accurate
results can be obtained for the magnetization and susceptibilities
in different  parameter
regions of the impurity Anderson model
with  strong electron correlation. 
 The RPT approach is a physically transparent one where 
we follow the flow equations for the renormalization of the quasiparticles
from their almost bare values
in a weak correlation regime to  strongly renormalized ones in the strong
correlation regime.
 In contrast to other renormalization techniques which involve
eliminating or integrating out of higher energy states, such as the NRG,
we need to consider  the flow of a restricted number of parameters only, those  required to specify the low energy behavior,
such as  $\tilde \Delta$ and
$\tilde h$, or  $\tilde \Delta$ and
$\tilde \epsilon_d$. This approach opens up the possibility of applications to a much wider class of models involving strong electron correlation.

\par 

\bigskip
 \noindent{\bf Acknowledgement}\par
 \bigskip\par
{We thank Johannes Bauer,   Daniel Crow, Nicolas Dupuis  for helpful
  discussions. KE and VP acknowledge the support of an EPSRC grant.
\par

\section{Appendix A}

 We consider the derivation of the result in Eqn. (\ref{w1}) where we change only the parameter
$\mu_1=h$. We can use the Friedel sum rule,
\begin{eqnarray}
\lefteqn{n_{d,\sigma}(h+\delta h)={1\over 2}}\nonumber \\
&&-{1\over\pi}{\rm
  tan}^{-1}\left({\epsilon_d-\sigma(h+\delta h)+\Sigma_\sigma(0,h+\delta
  h)\over\Delta}\right).\end{eqnarray}
Expanding this to first order in $\delta h$,
\begin{eqnarray}
\lefteqn{n_{d,\sigma}(h+\delta h)=n_{d,\sigma}(h)}\nonumber\\
&&+\sigma \delta h \rho_\sigma(0,h)
\left(1-\sigma{\partial \Sigma_\sigma(h)\over\partial h}\right) +{\rm O}((\delta
h)^2),\label{q1}\end{eqnarray}
where
\begin{equation}\rho_\sigma(\omega,h)={1\over\pi}{\Delta\over (\omega-\epsilon_d+\sigma h
-\Sigma_\sigma(0,h))^2+\Delta^2}.\end{equation}
We also have a Ward identity in the form,
\begin{equation}{\partial\Sigma_\sigma(\omega,h)\over\partial \omega}|_{\omega=0}
-\sigma{\partial\Sigma_\sigma(0,h)\over\partial h}=\rho_{-\sigma}
(0,h)\Gamma_{\uparrow,\downarrow}(0,0,0,0),\end{equation}
which can be rewritten in the form, 
\begin{equation}1-
\sigma{\partial\Sigma_\sigma(0,h)\over\partial h}=\rho_{-\sigma}
(0,h)\Gamma_{\sigma,-\sigma}(0,0,0,0)+{1\over z_\sigma}.\end{equation}
Using this result in Eqn. (\ref{q2}) we obtain
\begin{eqnarray}
\lefteqn{n_{d,\sigma}(h+\delta h)=n_{d,\sigma}(h)}\nonumber\\
&&+\sigma \delta h
\tilde\rho_\sigma(0,h)(1+\tilde U(h)\tilde\rho_{-\sigma}(0,h)) +{\rm O}((\delta
h)^2),\end{eqnarray}
where we have used the fact the $\tilde\rho_\sigma(0,h)=\rho_\sigma(0,h)/z_\sigma$.\par
We also have
\begin{eqnarray}
\lefteqn{n_{d,\sigma}(h+\delta h)={1\over 2}}\nonumber\\
&&-{1\over\pi}{\rm
  tan}^{-1}\left({\tilde\epsilon_d(h)-\sigma\tilde h(h)+\tilde \Sigma_\sigma(0,h,\delta
  h)\over\tilde\Delta_{\sigma}(h)}\right).\end{eqnarray}
Expanding this to first order in $\delta h$,
\begin{eqnarray}
\lefteqn{n_{d,\sigma}(h+\delta h)=n_{d,\sigma}(h)}\nonumber\\
&&-\sigma \delta h\tilde
\rho_\sigma(0,h)\tilde\Sigma_\sigma(0,h,\delta h)+{\rm O}((\delta
h)^2),\end{eqnarray}
Equating the two expressions we find
\begin{eqnarray}
\lefteqn{\alpha_{1,\sigma}(h,\delta  h)= \tilde \Sigma_\sigma(0,h,\delta
  h)=}\nonumber\\
&&-\sigma(1+\tilde U(h)\tilde\rho_{-\sigma}(0,h))\delta h+{\rm O}((\delta
h)^2).\end{eqnarray}

Using a very similar argument we can derive the result in Eqn. (\ref{w2})
  where we consider a change
  $\delta{\mu_2}=\epsilon_d$.
\begin{equation}
n_{d,\sigma}(\epsilon_d+\delta \epsilon_d)={1\over 2}-{1\over\pi}{\rm
  tan}^{-1}\left({\epsilon_d+\delta\epsilon_d+\Sigma_\sigma(0,\epsilon_d+\delta
  \epsilon_d)\over\Delta}\right).\end{equation}
Expanding this to first order in $\delta \epsilon_d$,
\begin{eqnarray}
\lefteqn{n_{d,\sigma}(\epsilon_d+\delta
  \epsilon_d)=n_{d,\sigma}(\epsilon_d)}\nonumber\\
&&-\delta \epsilon_d \rho_\sigma(0,\epsilon_d)
\left(1+{\partial \Sigma_\sigma(\epsilon_d)\over\partial \epsilon_d}\right) +{\rm O}((\delta
\epsilon_d)^2).\label{q2}\end{eqnarray}
We also have a Ward identity in the form,
\begin{equation}{\partial\Sigma_\sigma(\omega,h)\over\partial \omega}|_{\omega=0}+
{\partial\Sigma_\sigma(0,\epsilon_d)\over\partial \epsilon_d}=-\rho_{-\sigma}
(0,h)\Gamma_{\sigma,-\sigma}(0,0,0,0),\end{equation}
which can be rewritten in the form, 
\begin{equation}1+
{\partial\Sigma_\sigma(0,\epsilon_d)\over\partial \epsilon_d}=-\rho_{-\sigma}
(0,\epsilon_d)\Gamma_{\sigma,-\sigma}(0,0,0,0)+{1\over z_\sigma}.\end{equation}
Substituting into Eqn. (\ref{q2}) we find
\begin{eqnarray}
\lefteqn{n_{d,\sigma}y(\epsilon_d+\delta
  \epsilon_d)=n_{d,\sigma}(\epsilon_d)}\nonumber\\
&&-\delta \epsilon_d
\tilde \rho_\sigma(0,\epsilon_d)(1-\tilde U(\epsilon_d)\tilde\rho_{-\sigma}(0,\epsilon_d)) +{\rm O}((
\delta \epsilon_d)^2),\end{eqnarray}
We also have
\begin{equation}
n_{d,\sigma}(\epsilon_d+\delta \epsilon_d)={1\over 2}-{1\over\pi}{\rm
  tan}^{-1}\left({\tilde\epsilon_d+\tilde \Sigma_\sigma(0,\epsilon_d,\delta
  \epsilon_d)\over\tilde \Delta(\epsilon_d)}\right).\end{equation}
Expanding this to first order in $\delta \epsilon_d$,
\begin{equation}
n_{d,\sigma}(\epsilon_d+\delta \epsilon_d)=n_{d,\sigma}(\epsilon_d)-
\rho_\sigma(0,\epsilon_d)\tilde\Sigma_\sigma(0,\epsilon_d,\delta \epsilon_d)+{\rm O}((\delta
\epsilon_d)^2),\end{equation}
Equating the two expressions we find
\begin{eqnarray}
\lefteqn{\alpha_{1,\sigma}(\epsilon_d,\delta
  \epsilon_d)= \tilde \Sigma_\sigma(0,\epsilon_d,\delta
  \epsilon_d)=}\nonumber\\
&&(1-\tilde U(\epsilon_d)\tilde\rho_{-\sigma}(0,\epsilon_d))\delta \epsilon_d+{\rm O}((\delta
\epsilon_d)^2).\end{eqnarray}

\section{Appendix B}

For the model with particle-hole symmetry the  dynamic transverse spin
susceptibility for the free quasiparticles (particle with spin $\uparrow$ and
the
hole with spin $\downarrow$)
in a magnetic field is given by
\begin{equation}
\tilde \Pi^t_{ph}(\omega)={i\over \pi}(F^{ph}+
K^{ph})
\end{equation}
for $\omega>0$, where
\begin{equation}
F^{ph}(\omega)=-{1\over
  \omega+2\tilde h}\,{\rm
  ln}\left[{\omega+\tilde h+i\tilde\Delta\over
  -\tilde h+i\tilde\Delta}\right]
\end{equation}

\begin{equation}
K^{ph}(\omega)={1\over
  \omega+2\tilde h+2i\tilde\Delta}\,{\rm
  ln}\left[{\omega+\tilde h+i\tilde\Delta\over
  \tilde h+i\tilde\Delta}\right]
\end{equation}
 The causal propagator $\tilde \Pi_{ph}(\omega)$ is an even function of
 $\omega$ in the absence of a magnetic field.\par
The two-particle propagator for free quasiparticles  $\tilde \Pi_{pp}(\omega)$
is given by
\begin{equation}
\tilde \Pi_{pp}(\omega)={i\over 2\pi}(F^{pp}_{\sigma,-\sigma}+F^{pp}_{-\sigma,\sigma}+
K^{pp}_{\sigma,-\sigma}+K^{pp}_{-\sigma,\sigma})
\end{equation}
for $\omega>0$, where
\begin{equation}
F^{pp}_{\sigma,-\sigma}(\omega)=-{1\over
  \omega-2\tilde\epsilon_d+i\tilde\Delta_\sigma -i\tilde\Delta_{-\sigma}}{\rm
  ln}\left[{\omega-\tilde\epsilon_{d,\sigma}+i\tilde\Delta_\sigma\over
  \tilde\epsilon_{d,-\sigma}+i\tilde\Delta_{-\sigma}}\right]
\end{equation}
and
\begin{equation}
K^{pp}_{\sigma,-\sigma}(\omega)={1\over
  \omega-2\tilde\epsilon_d+i\tilde\Delta_\sigma +i\tilde\Delta_{-\sigma}}{\rm
  ln}\left[{\omega-\tilde\epsilon_{d,\sigma}+i\tilde\Delta_\sigma\over
  -\tilde\epsilon_{d,\sigma}+i\tilde\Delta_\sigma}\right]
\end{equation}
 In the absence of a magnetic field the causal propagator $\tilde \Pi_{pp}(\omega)$ is an even function of $\omega$.

%\bibliography{artikel}

\bibliographystyle{h-physrev3}
\end{document}